\documentclass[aip,jap,reprint,superscriptaddress]{revtex4-1}

\usepackage{graphicx}
\usepackage{dcolumn}
\usepackage{amsmath}
\usepackage{textcomp}

\begin{document}

\title{The non-equilibrium response of a superconductor to pair-breaking radiation measured over a broad frequency band} 

\author{P.J. de Visser}
\email{p.j.devisser@tudelft.nl}
\altaffiliation{Current address: Department of Quantum Matter Physics, University of Geneva, Geneva 1211, Switzerland}
\affiliation{Kavli Institute of NanoScience, Faculty of Applied Sciences, Delft University of Technology, Lorentzweg 1, 2628 CJ Delft, The Netherlands}

\author{S.J.C. Yates}
\affiliation{SRON Netherlands Institute for Space Research, Landleven 12, 9747AD Groningen, The Netherlands}

\author{T. Guruswamy}
\affiliation{Quantum Sensors Group, Cavendish Laboratory, University of Cambridge, JJ Thomson Avenue, Cambridge CB3 0HE, United Kingdom}

\author{D.J. Goldie}
\affiliation{Quantum Sensors Group, Cavendish Laboratory, University of Cambridge, JJ Thomson Avenue, Cambridge CB3 0HE, United Kingdom}

\author{S. Withington}
\affiliation{Quantum Sensors Group, Cavendish Laboratory, University of Cambridge, JJ Thomson Avenue, Cambridge CB3 0HE, United Kingdom}

\author{A. Neto}
\affiliation{Terahertz Sensing Group, Faculty of Electrical Engineering, Mathematics and Computer Science, Delft University of Technology, Mekelweg 4, 2628 CD Delft, The Netherlands}

\author{N. Llombart}
\affiliation{Terahertz Sensing Group, Faculty of Electrical Engineering, Mathematics and Computer Science, Delft University of Technology, Mekelweg 4, 2628 CD Delft, The Netherlands}

\author{A.M. Baryshev}
\affiliation{SRON Netherlands Institute for Space Research, Landleven 12, 9747AD Groningen, The Netherlands}
\affiliation{Kapteyn Astronomical Institute, University of Groningen, Landleven 12, 9747 AD Groningen, The Netherlands}

\author{T.M. Klapwijk}
\affiliation{Kavli Institute of NanoScience, Faculty of Applied Sciences, Delft University of Technology, Lorentzweg 1, 2628 CJ Delft, The Netherlands}
\affiliation{Physics Department, Moscow State Pedagogical University, Moscow 119991, Russia}

\author{J.J.A. Baselmans}
\affiliation{SRON Netherlands Institute for Space Research, Sorbonnelaan 2, 3584 CA Utrecht, The Netherlands}
\affiliation{Terahertz Sensing Group, Faculty of Electrical Engineering, Mathematics and Computer Science, Delft University of Technology, Mekelweg 4, 2628 CD Delft, The Netherlands}

\date{\today}

\begin{abstract}

We have measured the absorption of terahertz radiation in a BCS superconductor over a broad range of frequencies from 200 GHz to 1.1 THz, using a broadband antenna-lens system and a tantalum microwave resonator. From low frequencies, the response of the resonator rises rapidly to a maximum at the gap edge of the superconductor. From there on the response drops to half the maximum response at twice the pair-breaking energy. At higher frequencies, the response rises again due to trapping of pair-breaking phonons in the superconductor. In practice this is the first measurement of the frequency dependence of the quasiparticle creation efficiency due to pair-breaking in a superconductor. The efficiency, calculated from the different non-equilibrium quasiparticle distribution functions at each frequency, is in agreement with the measurements. 

\end{abstract}

\maketitle

In a superconductor at low temperature, most of the electrons are bound in Cooper pairs. These pairs can be broken into quasiparticles by absorbing photons with an energy larger than the binding energy. This mechanism is frequently used to detect submillimetre and terahertz radiation using conventional superconductors such as aluminium. Pair-breaking detectors are usually assumed to measure the number of quasiparticles created by the absorbed radiation. The observable that measures the number of quasiparticles varies from the complex conductivity for microwave kinetic inductance detectors \cite{pday2003} (MKIDs), the current through a tunnel junction \cite{apeacock1996} to the capacitance of a small superconducting island \cite{pechternach2013}. These observables are mainly sensitive to quasiparticles with an energy close to the gap energy of the superconductor, $\Delta$. The working principle of these detectors is usually explained in terms of an effective number of quasiparticles which is maintained by a balance between the radiation power and electron-phonon interaction (recombination) \cite{arothwarf1967}. To convert the power ($P$) into a number of quasiparticles ($N_{qp}$), the quasiparticle creation efficiency $\eta_{pb}$ is introduced, which compares the actual $N_{qp}$ with the maximum possible $N_{qp}$ when all created quasiparticles would have an energy $\Delta$. Since Cooper pairs have a binding energy of $2\Delta$, a photon with an energy in between $2\Delta$ and $4\Delta$ can still only create two quasiparticles. The rest of the energy is lost through electron-phonon scattering, hence $\eta_{pb}<1$. For higher energies $\eta_{pb}$ depends on the phonon trapping factor, which determines whether high energy phonons are directly lost or can break an additional pair. $\eta_{pb}$ is therefore not an efficiency in the sense that photons are lost, but it reduces the detector responsivity. MKIDs are superconducting microwave resonators which sense the number of quasiparticles through the complex conductivity of the superconductor. The phase response ($\theta$) of such a resonator can be approximated by
\begin{equation}
 \theta \propto -\frac{\delta\sigma_2}{\sigma_2} \propto \delta N_{qp} \propto \eta_{pb}P,
\label{eq:responseNqp}
\end{equation}
where $\sigma_2$ is the imaginary part of the complex conductivity. For the last proportionality we assume to be in the linear regime where the quasiparticle recombination lifetime does not change significantly upon a change in $N_{qp}$. $N_{qp}$ is dominated by background power and $\delta N_{qp}\ll N_{qp}$. $\eta_{pb}$, and hence the detector response, is dependent on the frequency of the absorbed photons, even at constant absorbed power.

On a microscopic level, the pair-breaking radiation leads to injection of quasiparticles at very specific energies \cite{bivlev1973}. Together with electron-phonon interaction (scattering and recombination) \cite{jchang1978} a non-equilibrium, non-thermal quasiparticle energy distribution $f(E)$ is formed, which determines the response to pair-breaking radiation as recently shown in Ref. \onlinecite{tguruswamy2014}. For microwave resonators this is reflected in the explicit dependence of $\sigma_2$ on $f(E)$ \cite{dmattis1958}:
\begin{eqnarray}
\frac{\sigma_2}{\sigma_N} &=& \frac{1}{\hbar \omega}\int^{\Delta}_{\Delta-\hbar\omega}[1-2f(E+\hbar\omega)]g_2(E)dE,
\label{eq:sigmatwo}\\
g_2(E) &=& \frac{E^2+\Delta^2+\hbar\omega E}{(\Delta^2-E^2)^{1/2}[(E+\hbar\omega)^2-\Delta^2]^{1/2}}, 
\label{eq:g1g2f}
\end{eqnarray}
where $\sigma_N$ is the normal state conductivity, $\hbar$ Planck's constant and $\omega$ the microwave frequency. $\eta_{pb}$ is thus an attempt to capture all information contained in $f(E)$ in a single number, to allow for an effective quasiparticle number approach as given by Eq. \ref{eq:responseNqp}. 

Here we present the first measurement of $\eta_{pb}$ over a broad range in frequencies close to the superconducting gap (350 - 1100 GHz). A Ta MKID is used as the detector in a Fourier transform spectrometer (FTS) to measure the frequency dependence of the response. The measured response curve of the detector can be well explained by a frequency dependent $\eta_{pb}$, caused by a different non-equilibrium $f(E)$ calculated for different pair-breaking frequencies.

From an applied point of view MKIDs \cite{jzmuidzinas2012} are considered promising detectors for large arrays due to the intrinsic ease of multiplexing their readout. MKIDs are photon noise limited for various frequencies \cite{syates2011,rjanssen2013,rjanssen2014,pdevisser2014,pmauskopf2014,jhubmayr2015}. The level of experimental detail that has now been achieved \cite{pdevisser2014,pdevisser2014b,rjanssen2014b} calls for a more detailed understanding of the absorption of radiation. An important gap in this understanding is a measurement of $\eta_{pb}$. $\eta_{pb}$ determines key parameters: the responsivity of the detector, the recombination noise level in the photon-noise limited regime \cite{syates2011} and the sensitivity in the generation-recombination noise dominated limit \cite{pdevisser2011}. The common number used for $\eta_{pb}$ is 0.57 for all signal frequencies, which was derived for the temporal relaxation of very high energy excitations which first create a photo-electron \cite{mkurakado1982,akozorezov2000}, an approach which is not applicable for frequencies close to the gap.

\begin{figure}
\includegraphics[width=0.99\columnwidth]{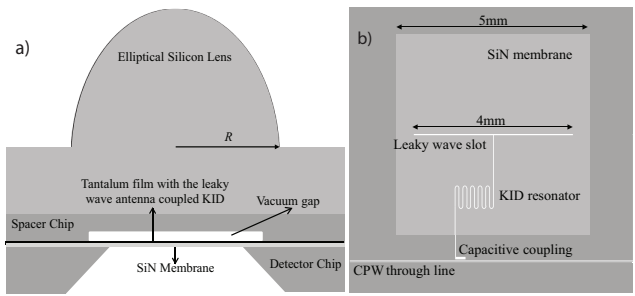}
\caption{\label{fig:leaky} (a) Schematic of the detector. The detector chip is fabricated on a SiN membrane and glued to an elliptical lens, leaving a small vacuum gap between the antenna and the lens-dielectric. (b) The design of the detector chip. The antenna slot, coupled to a microwave resonator, is fabricated on a SiN membrane (light grey square). The resonator is capacitively coupled to a microwave readout line. }
\end{figure}

Previous studies of the absorption of radiation in superconductors have either measured $f(E)$ directly with tunnel-junctions \cite{fjaworski1979,asmith1980}, but only with a single-frequency optical laser, or measured the absorption over a broad band with a bolometer \cite{kkornelson1991,mdressel2013}, which is insensitive to the non-equilibrium effects that determine $\eta_{pb}$. To measure $\eta_{pb}$ over a broad frequency band, a known and relatively constant radiation power over a broad frequency band is required. Secondly, we need the absorption of all of that power at all frequencies within the volume of the detector to exclude the effect of frequency dependent absorption \cite{kkornelson1991}. We therefore use a particularly wideband lens-antenna system, which is based on the leaky-wave antenna \cite{aneto2010a} and shown in Fig. \ref{fig:leaky}. It consists of a 30 $\mu$m wide, 4 mm long slot, etched in a 200 nm thick Ta film with a resistivity of $6.7$ $\mu\Omega$cm, which is sputter deposited onto a 3 $\mu$m thick SiN membrane using a 6 nm Nb seed layer. A spacer chip, placed in between the Ta and the Si lens ensures a 35 $\mu$m vacuum gap between the metal layer and the Si lens, which is crucial to get a high directivity of the antenna over a broad frequency band \cite{aneto2010a}. This lens-antenna was demonstrated to have very clean beampatterns over the frequency range 300-900 GHz. The antenna launches the signal as a travelling wave into a coplanar waveguide (CPW) with a central strip of 3.5 $\mu$m and slots of 3 $\mu$m wide, which length is designed to make a quarter wavelength resonator at 4.6571 GHz (the MKID detector). An extensive discussion of the design, fabrication process and beampattern measurements can be found in Ref. \onlinecite{aneto2014}.

\begin{figure}
\includegraphics[width=0.99\columnwidth]{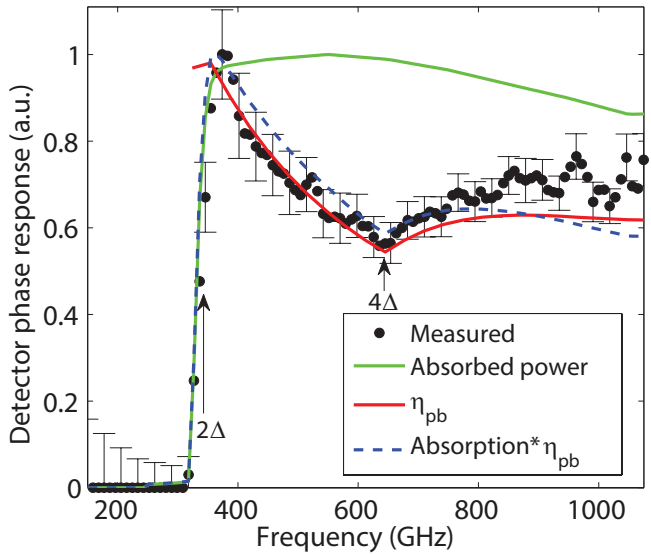}
\caption{\label{fig:measuredFTS} The measured phase response (dots) of the microwave resonator as a function of the frequency of the pair-breaking radiation and normalised to one. Error bars are shown every third point. The green line represents a calculation of the power absorption of the superconducting transmission line. The red line is a simulation of the pair-breaking efficiency (not normalised) that arises due to the different quasiparticle distributions at different excitation frequencies. The blue dashed line combines the two effects. The blue and green lines are both normalised to one.}
\end{figure}

The detector is cooled down in a $^{3}$He/$^{4}$He cryostat to a bath temperature of 320 mK. The cryostat has optical access through a window, Goretex infrared blockers at 77 K and 4 K, a 1.1 THz lowpass filter at 4 K. The Michelson Fourier transform spectrometer (FTS) consists of a globar source at 2000 \textcelsius, a fixed and a movable mirror and a mylar beamsplitter. To eliminate absorption lines due to water, the FTS is placed in vacuum. The MKID itself is the detector in this setup. The phase response of the detector was measured as a function of the mirror distance. The phase response is linear in power, which is verified using the response to a full rotation of a polariser in a separate measurement (i.e. the last proportionality in Eq. \ref{eq:responseNqp} is valid). The Fourier transform of the interferogram, corrected for the frequency dependence of the filters and beamsplitter (see Supplementary Figure S1 \footnote{See supplemental material at [URL will be inserted by AIP] for the setup corrections applied to the raw response.}), is shown in Fig. \ref{fig:measuredFTS} as black dots, which is the central result of this letter. The beamsplitter response contains a cross-polarisation contribution of 28$\pm$5 \%, which is derived by integrating the measured beampatterns of the antenna \cite{aneto2014} over the opening angle of the source (Supplementary Note 1). The other contribution to the error bars on the data is given by the uncertainty in the exact beamsplitter thickness $48\pm 2$ $\mu$m.

The power as a function of frequency that arrives at the detector waveguide input can be calculated using
\begin{equation}
 P(\nu) = \frac{c^2}{4\pi}\int_\Omega \frac{A(\nu)B(\nu,T_{BB})}{2\nu^2}C(\nu)G(\nu,\Omega) d\Omega,
\label{eq:power}
\end{equation}
with $c$ the speed of light, $\Omega$ the solid angle, $A(\nu)$ the transmission of optical elements (filters, beamsplitter), $B(\nu,T_{BB})$ the brightness of the source given by Planck's law, $C(\nu)$ the antenna efficiency and $G(\nu,\Omega)$ the antenna gain pattern. The factor $(c/\nu)^2$ reflects a single mode throughput. For the purpose of the present experiment it is sufficient to know the relative power at each frequency. As discussed in Ref. \onlinecite{aneto2014} the beam patterns are measured in three frequency windows: 290-350 GHz, 640-710 GHz and 790-910 GHz. The difference in the directivity for these bands is compensated by the difference in the part of the source that they capture. The brightness of the blackbody at the measured frequencies can be well described in the Rayleigh-Jeans limit, where $B(\nu,T_{BB}) = 2kT_{BB}\nu^2/c^2$, which exactly compensates the frequency dependence due to the throughput $(c/\nu)^2$. The antenna efficiency is the only element from Eq. \ref{eq:power} that introduces a frequency dependence, as shown in Supplementary Figure 2. Using Eq. \ref{eq:power} we estimate the absorbed pair-breaking power from the FTS to be 15 nW. The transmission of the optical elements $A(\nu)$ is taken into account in the correction of the measured response as explained above.

In Fig. \ref{fig:measuredFTS}, starting from 200 GHz, we observe no response until 320 GHz where the absorption rises drastically because photons have enough energy to break Cooper pairs ($2\Delta$). This steep rise in response is partially the well-known absorption edge of the superconductor \cite{rglover1957}: the frequency dependent absorption of a plain superconducting film through the complex sheet impedance. However in this experiment the antenna collects the radiation and launches it as a travelling wave into the MKID CPW. For frequencies well above the gap, it takes only 1 mm to absorb 90\% of the power, thus all power is absorbed in the detector volume. Therefore the non-monotonous sheet resistance for frequencies above the gap does not affect the measured response in this experiment, which is crucial to make the non-equilibrium response of the superconductor visible. The percentage of the power absorbed in the CPW line is calculated using the attenuation constant of a CPW \cite{cholloway1995,pdevisserphd} based on the frequency dependent complex conductivity of the Ta film following Mattis and Bardeen \cite{dmattis1958}. We assume the absorption length to be 10.4 mm, twice the length of the resonator. Radiation that is not absorbed (only for $h\nu<2\Delta$) will be reemitted by the antenna. The effective temperature of the superconductor is assumed to be 1.2 K, to be consistent with a linear response to 15 nW of FTS signal ($N_{qp}$ is dominated by the effective temperature). The resulting frequency dependent absorption is shown as the green line in Fig. \ref{fig:measuredFTS}. The maximum around 550 GHz is due to the simulated efficiency of the antenna, which is also taken into account (see Supplementary Fig. S4).

For frequencies higher than 400 GHz, the power received by the antenna is fully absorbed in the detector waveguide. However, in Fig. \ref{fig:measuredFTS}a we observe a drop in the response close to 650 GHz ($4\Delta$) by about a factor of two, after which the response increases again. Having taken into account all frequency dependent power contributions, the only parameter left is the frequency dependence of the non-equilibrium response of the superconductor, represented by $\eta_{pb}$ in Eq. \ref{eq:responseNqp}. 

\begin{figure}
\includegraphics[width=0.99\columnwidth]{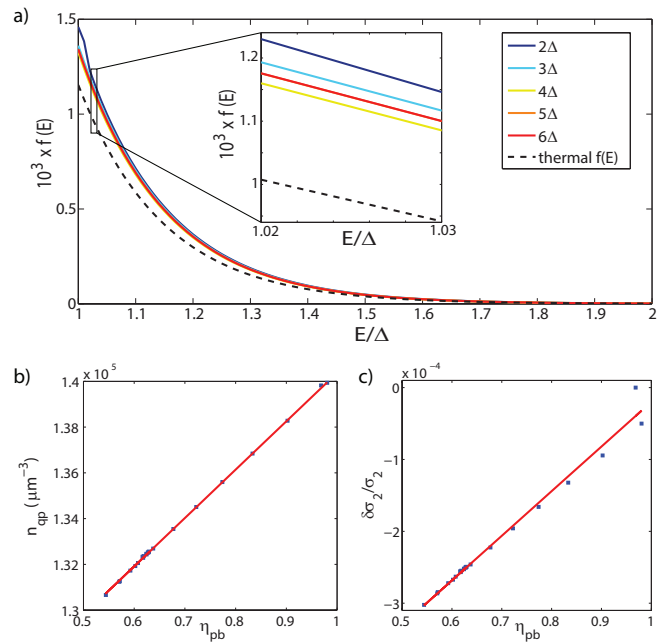}
\caption{\label{fig:distributionfunctions} (a) The calculated quasiparticle distribution functions $f(E)$ as a function of energy for different excitation energies (2$\Delta$, 3$\Delta$, 4$\Delta$, 5$\Delta$ and 6$\Delta$). The lines for $5\Delta$ and $6\Delta$ coincide. The absorbed power was kept constant, and the resulting variation is thus only an effect of the frequency of the absorbed photons. The inset highlights the differences in $f(E)$ on a small energy scale. (b) The change in the quasiparticle density ($n_{qp}$) and (c) the imaginary part of the complex conductivity $\delta\sigma_2/\sigma_2$ as a function of the pair breaking efficiency $\eta_{pb}$. The different points are calculated at different excitation frequencies, 320 - 1100 GHz. The red lines are linear fits to the simulated points}
\end{figure}

The non-equilibrium distribution of quasiparticles is calculated using a quasiparticle creation term that describes the probability of creating a quasiparticle at a certain energy by breaking a Cooper pair following Eliashberg \cite{geliashberg1970,bivlev1973}. In steady state, the injection of quasiparticles at that energy is balanced by electron-phonon interaction (scattering and recombination). The kinetic equations for the non-equilibrium quasiparticle- and phonon energy distributions are solved following the approach by Chang and Scalapino \cite{jchang1978}. The numerical procedure is explained in Ref. \onlinecite{dgoldie2013}. In the modelling we used an effective bath temperature of 1.2 K to account for the broadband power absorbed by the detector. It was not necessary for this temperature to be exact as $\eta_{pb}$ is not strongly dependent on the bath temperature at low reduced temperatures (here $T/T_c = 0.27$) \cite{tguruswamy2014}. The resulting distribution functions $f(E)$ \emph{for constant absorbed power} are shown for various frequencies in Fig. \ref{fig:distributionfunctions}a. For higher excitation frequencies there are more quasiparticles with a higher energy, and therefore less weight close to the gap, where the resonator is sensitive. We therefore expect the maximum resonator signal at $\nu=2\Delta/h$ and a minimum at $\nu=4\Delta/h$. The pair breaking efficiency, $\eta_{pb}$, calculated from these distribution functions is shown in Fig. \ref{fig:measuredFTS} (red line). Combined with the frequency dependent absorption (dashed blue line) it clearly describes the main shape of the measured response.

It was shown by Guruswamy et al. \cite{tguruswamy2014} that the behaviour of $f(E)$ for frequencies higher than $\nu=4\Delta/h$ crucially depends on the phonon trapping factor. When phonons are released due to scattering or recombination, the ratio of their escape time $\tau_{esc}$ and the pair-breaking time $\tau_{pb}$ determines how many quasiparticles can be generated from a single incoming photon. Only for $\tau_{esc}/\tau_{pb}>1$ can $\eta_{pb}$ increase at energies above $4\Delta$. $\tau_{pb}$ is material dependent and equals 2.3$\times 10^{-11}$ s for Ta \cite{skaplan1976} (2.8$\times 10^{-10}$ s for Al). For 200 nm Ta on Si we obtain $\tau_{esc}=2$ ns \cite{skaplan1979}, which gives a trapping factor of 87, which makes Ta a favourable choice (over e.g. Al) to experimentally address the effect of phonon trapping. It is difficult to estimate the precise trapping factor because the substrate is a relatively thin membrane and because of the Nb seed layer, but it is certainly large. In practice, $\eta_{pb}$ is the same for trapping factors of 15 and higher \cite{tguruswamy2014}. The minimum at $4\Delta$ due to phonon trapping, which we observe in Fig. \ref{fig:measuredFTS} qualitatively distinguishes the non-equilibrium response from other frequency dependent phenomena. 

When Cooper pairs are broken, the created high energy quasiparticles relax back to energies close to the gap on a timescale of 0.1-10 ns \cite{akozorezov2000}. The response can well be described by an effective number of quasiparticles $N_{qp}$ \cite{tguruswamy2014} using $\eta_{pb}$ as in Eq. \ref{eq:responseNqp}. From the calculated $f(E)$, we derive $\eta_{pb}$, $n_{qp}$ (quasiparticle density) and $\sigma_2$. The (almost) linear relationship between those properties (Eq. \ref{eq:responseNqp}) is demonstrated in Figures. \ref{fig:distributionfunctions}b and c. We emphasise that although these figures and Eq. \ref{eq:responseNqp} suggest a simple effective $N_{qp}$, the knowledge of the microscopic $f(E)$ is needed to get that $N_{qp}$ at a certain ($P,\nu$) through $\eta_{pb}$. An effective $N_{qp}$ or effective temperature model \cite{jgao2008c,gcatelani2010} which may well describe a measurement at one specific frequency, even of the full $f(E)$ \cite{asmith1980}, would give constant $N_{qp}$ for constant absorbed power, regardless of the excitation frequency. It would therefore not explain our observations in Fig. \ref{fig:measuredFTS}. 

The qualitative agreement between measurement and simulation in Fig. \ref{fig:measuredFTS} is very good, especially the peak around $2\Delta$ and the characteristic $4\Delta$ point are well represented. The deviation that occurs at higher frequencies is most likely due to an incomplete understanding of the combination of the FTS system with the lens-antenna. Except for the mentioned uncertainties, the antenna efficiency and absorption length are not independently measured and the removal of the residual ripple in the response is not exact. To get a deviation smaller than the 10-15\% achieved now, one would need a complicated calibration with a bolometer with better sensitivity than the Ta MKID coupled to the same lens-antenna. We note that the characteristic impedance of the waveguide ($Z_0$) is also frequency dependent, but it changes the power transmitted from antenna to waveguide by only 0.1\%. The diffusion length of quasiparticles in Ta is 2 $\mu$m based on a recombination time of 50 ns (for an effective temperature of 1.2 K) and a diffusion constant of 0.8 cm$^2$s$^{-1}$ (Refs. \onlinecite{sfriedrich1997,tnussbaumer2000} and the measured resistivity). The effect of diffusion of quasiparticles from the central strip at the antenna feed is therefore negligible. Furthermore we checked that for this geometry radiation losses are a factor 10 lower than absorption in the superconductor \cite{mfrankel1991}.

The measured energy gap in the FTS response occurs at 324 GHz, corresponding to a $T_c$ of 4.4 K, assuming $2\Delta=3.52 k_B T_c$. This is consistent with the minimum response in Fig. \ref{fig:measuredFTS} at around 650 GHz ($4\Delta$). However the DC-measured $T_c$ of this film is 4.77 K, although most of our previous Ta films have also shown a $T_c$ of 4.4 K \cite{rbarends2009}. We presume that the Nb seed layer is thicker than anticipated giving a thin layer with a somewhat higher $T_c$ dominating the DC transport, whereas the radiation absorption is dominated by the lower gap in the thick Ta top layer.

The microwave readout power can strongly affect the response of a microwave resonator \cite{dgoldie2013,pdevisser2014b}. In this experiment we can neglect effects due to the absorbed readout power (1.8 nW), which is much smaller than the absorbed pair-breaking signal (15 nW). Readout power effects are only expected in the opposite limit \cite{tguruswamy2015}, which is nevertheless important to investigate in the future. The observed agreement of the measurements with the model is encouraging. At the same time it underlines the importance of understanding and controlling these parameters to optimise superconducting detectors.

We would like to thank Jan Barkhof for help with the FTS calibration. This work was in part supported by ERC starting Grant ERC-2009-StG Grant 240602 TFPA. T. M. Klapwijk acknowledges financial support from the Ministry of Science and Education of Russia under Contract No. 14.B25.31.0007 and from the European Research Council Advanced Grant No. 339306 (METIQUM). P. J. de Visser acknowledges support from a Niels Stensen Fellowship.

%

\clearpage
\onecolumngrid
\setcounter{figure}{0}

\renewcommand{\thefigure}{S\arabic{figure}}

\section*{Supplementary information for: 'The non-equilibrium response of a superconductor to pair-breaking radiation measured over a broad frequency band'} 

\subsection*{P.J. de Visser, S.J.C. Yates, T. Guruswamy, D.J. Goldie, S. Withington, A. Neto, N. Llombart, A.M. Baryshev, T.M. Klapwijk, and J.J.A. Baselmans}


\maketitle

\begin{figure*}[h]
\includegraphics[width=0.4\textwidth]{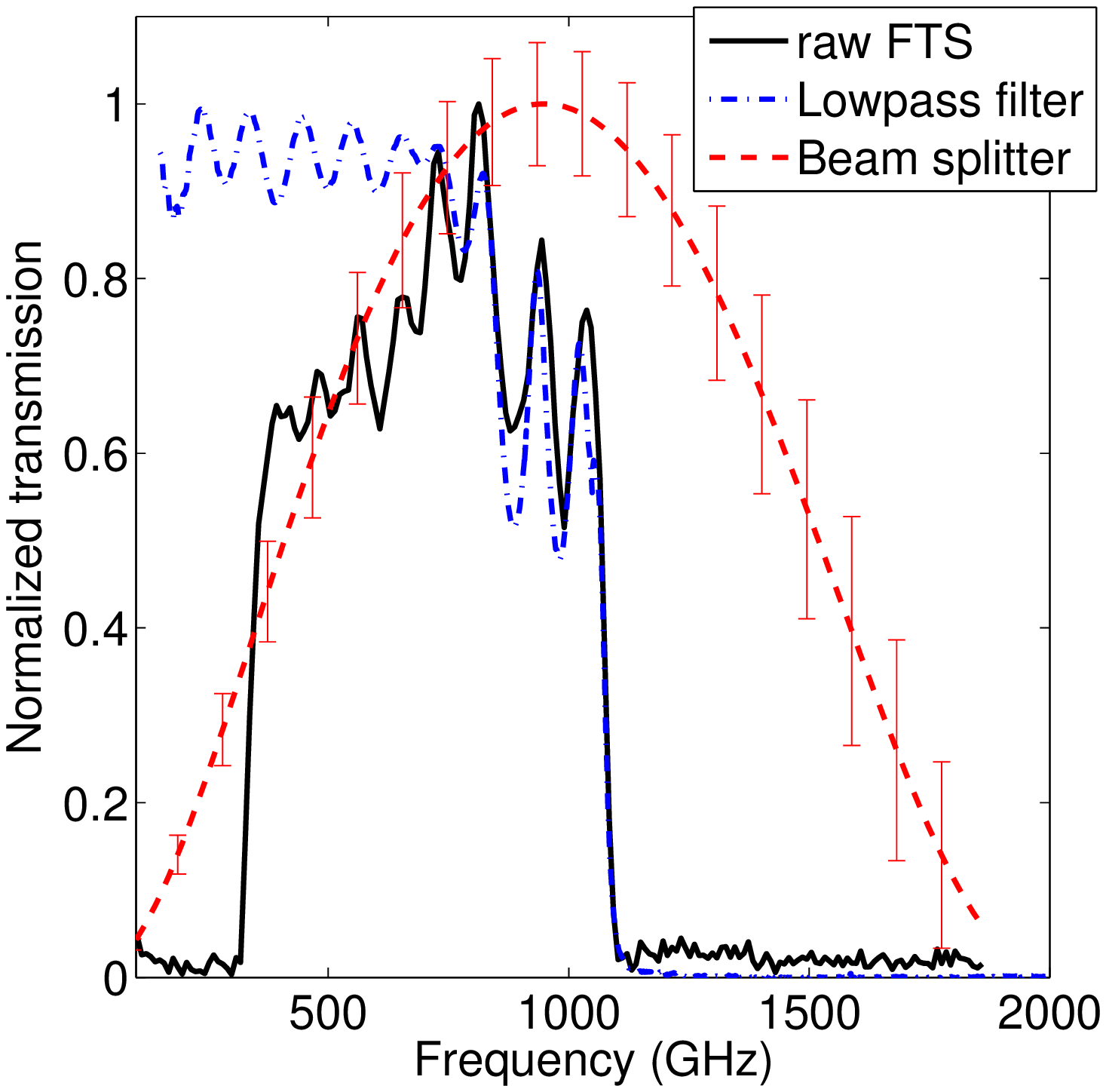}
\includegraphics[width=0.4\textwidth]{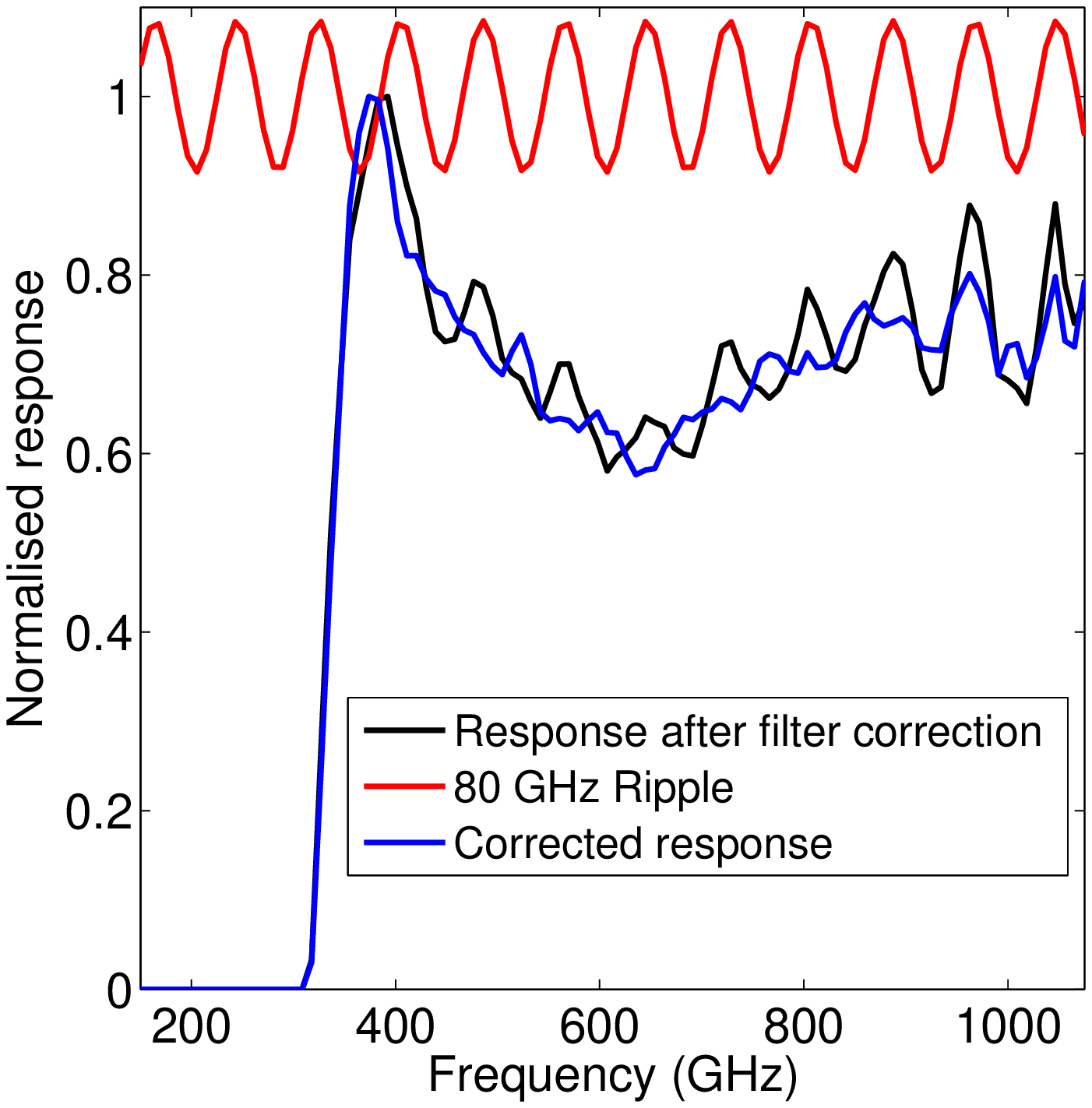}
\caption{\label{fig:corrections} (left) The raw FTS response as a function of frequency together with the applied corrections for the transmission of the beamsplitter (both p- and s-polarisation contributions) and the low-pass filter. The error bars on the calculated beam splitter response are due to an uncertainty in the beam splitter thickness induced by the affixation of the beam splitter to a metal frame. (right) The corrected response from the left panel, shown here as the black line, contains a residual setup ripple due to the source, which we correct for as shown by the red line. The resulting response (blue line) is the data shown in Fig. 2 of the main article.}
\end{figure*}

\section{Calculation of the cross-polarisation contribution to the FTS response}
\begin{figure*}
\includegraphics[width=0.5\textwidth]{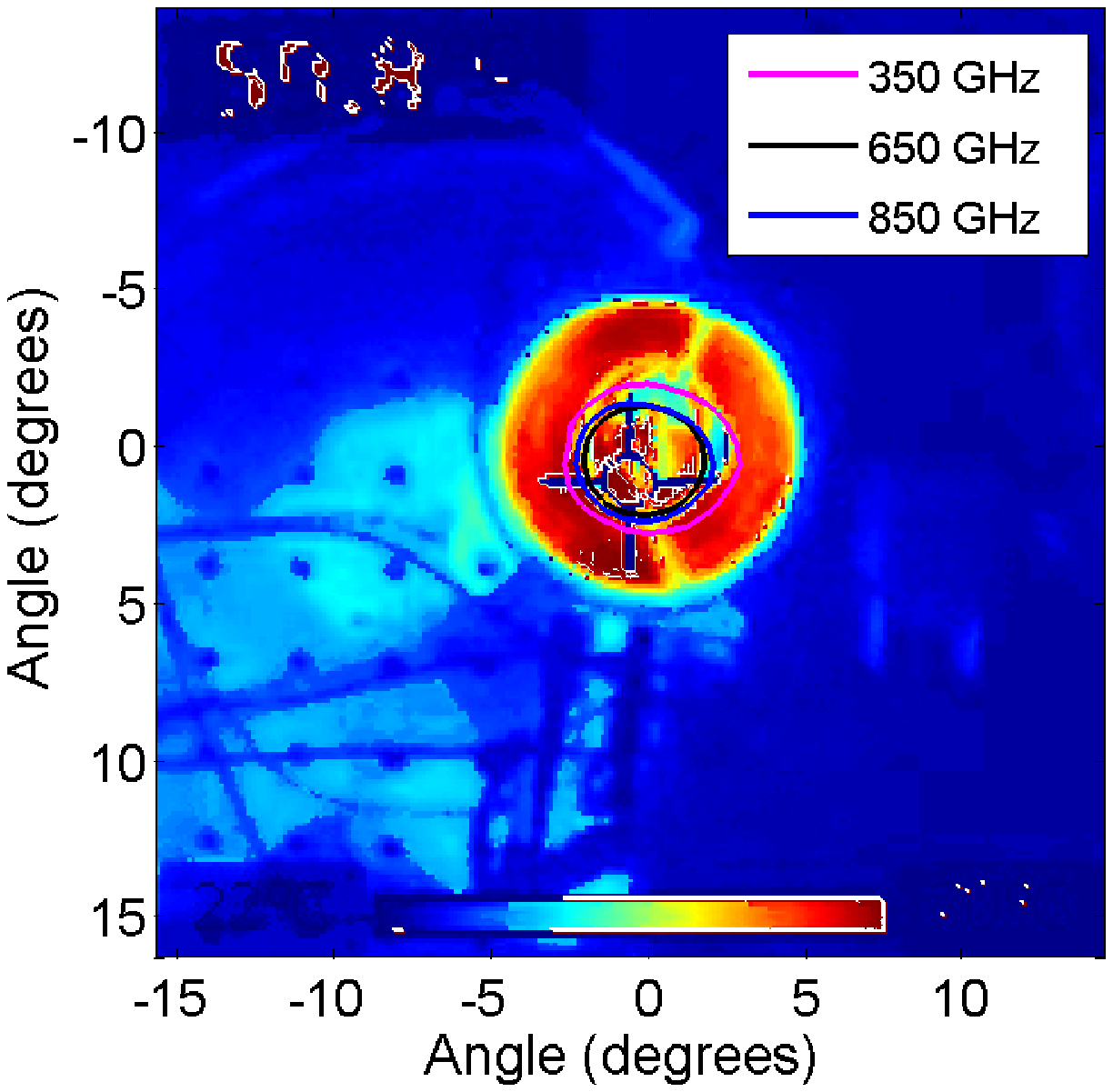}
\caption{\label{fig:IRsource} A photograph of the globar source taken with an infrared camera, demonstrating the 4.5 degree opening angle of the source. The solid lines in the centre are the 3 dB contours of the measured co-polarisation beampatterns at the frequencies indicated in the legend.}
\end{figure*}

\begin{figure*}
\includegraphics[width=0.32\textwidth]{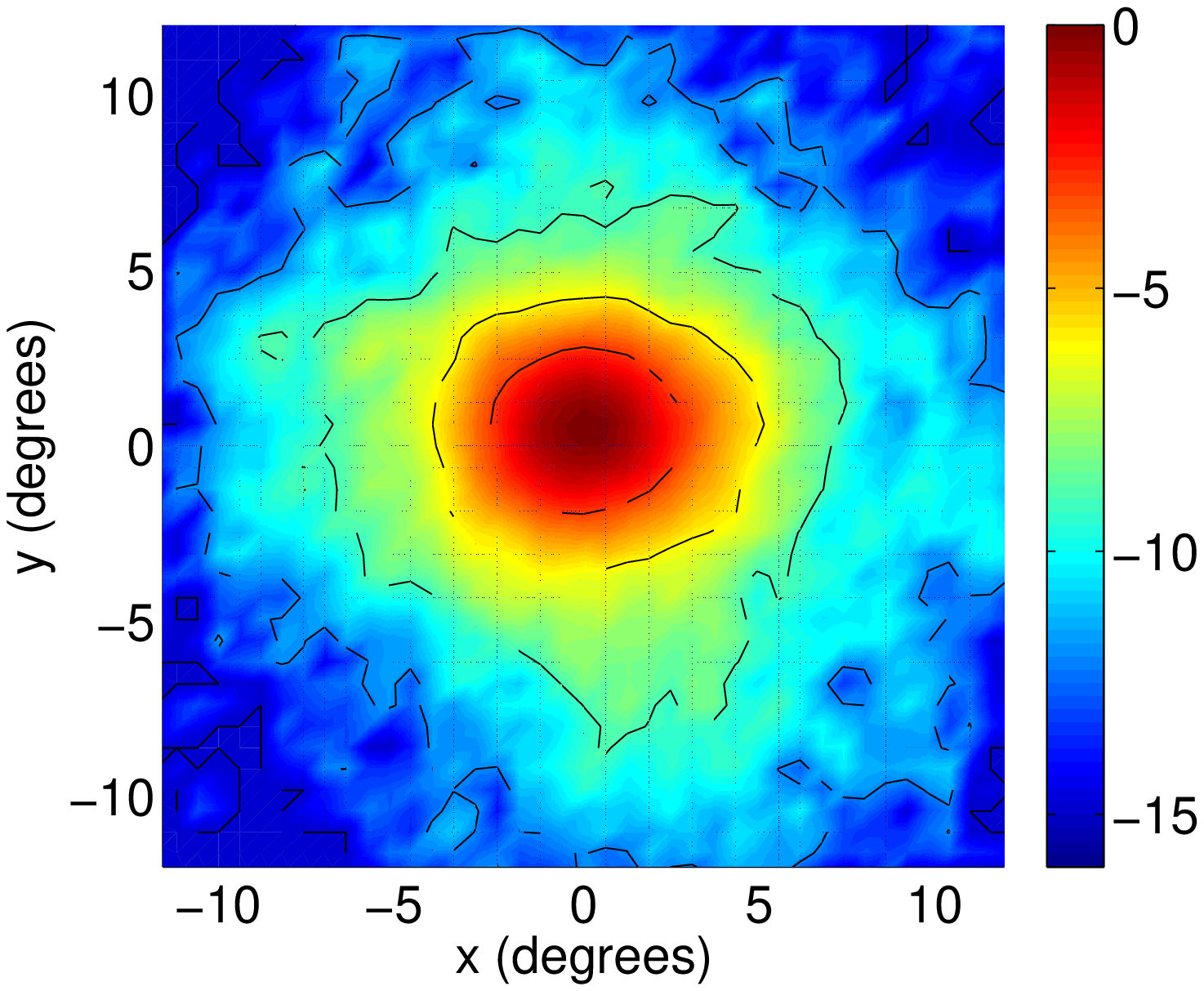}
\includegraphics[width=0.32\textwidth]{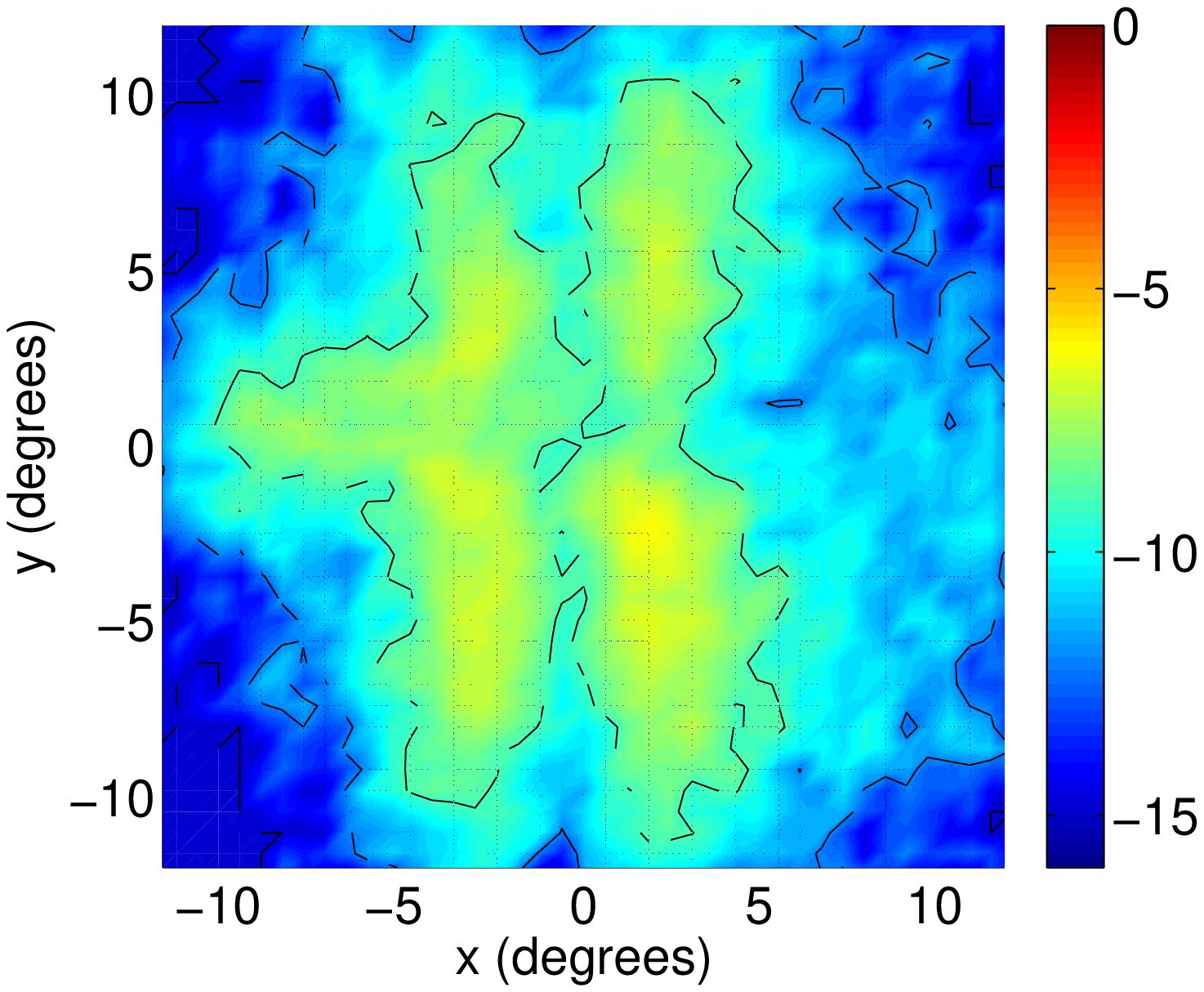}
\includegraphics[width=0.32\textwidth]{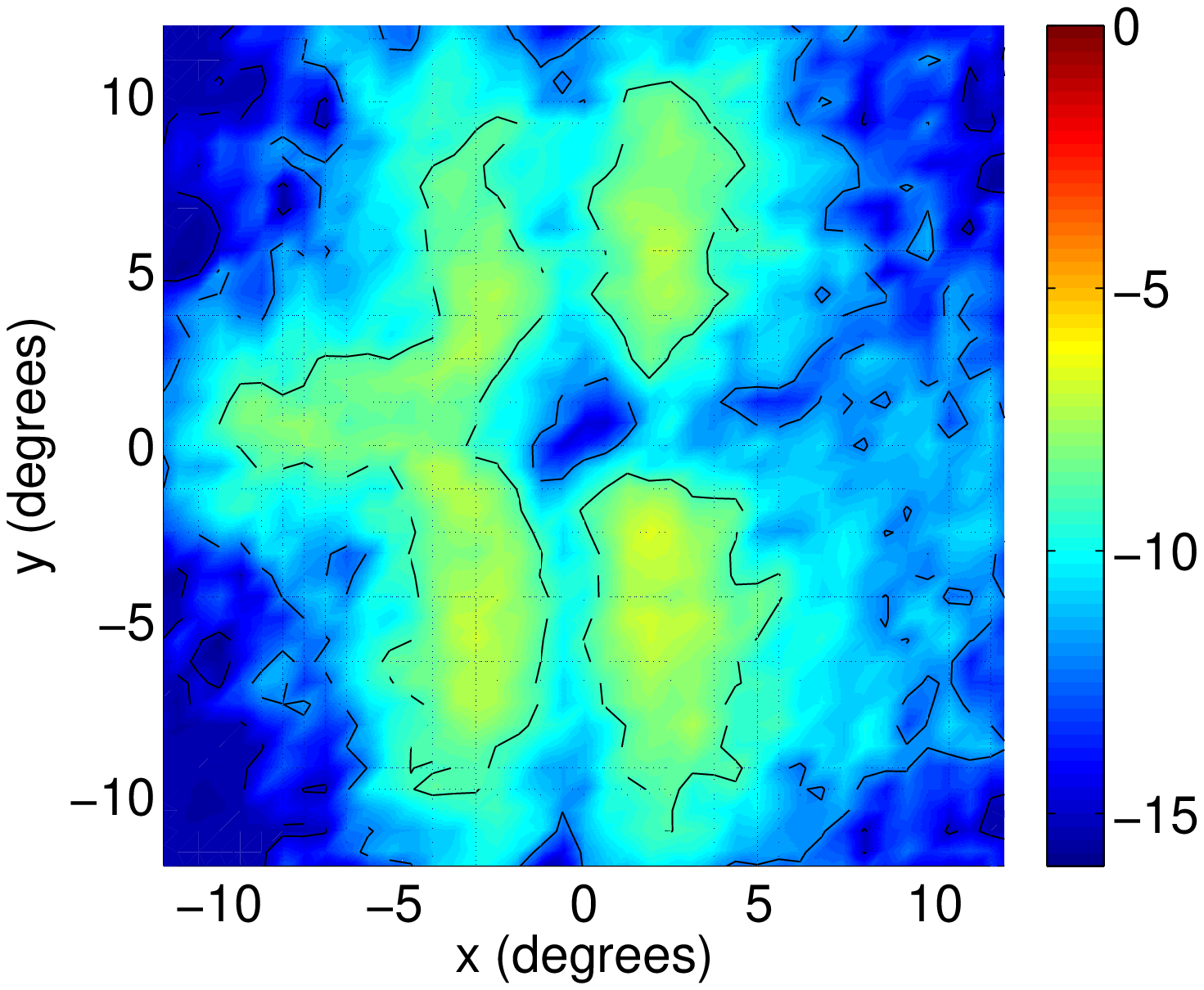}
\caption{\label{fig:beams350} (left) Measured beampattern in co-polarisation in a narrow band around 350 GHz. (middle) Measured cross-polarisation pattern. (right) Measured cross-polarisation pattern with a contribution of 9\% co-polarisation contribution subtracted. The color scales represent dB.}
\end{figure*}

The FTS response was measured without polarisers, which leaves the possibility of a cross-polar contribution in the response of the antenna. The beampatterns for co- and cross-polarisation were measured and reported in Ref. \onlinecite{aneto2014}. As an example we show the measured beam patterns for 350 GHz in Fig. \ref{fig:beams350}. We correct the cross-polar response for a small misalignment of the polariser by subtracting a 9\% co-polar contribution, the result of which is shown in the right panel of Fig. \ref{fig:beams350}. To calculate the cross/co-polar power ratio we first subtract the baseline level (-15 dB, due to noise) from all patterns. We subsequently integrate the beampatterns over an angle of $\pm$4.5$^{\circ}$, which corresponds to the opening angle of the broadband globar source as shown in Fig. \ref{fig:IRsource}. The resulting cross/co-polar response ratio is 0.27 for 350 GHz, 0.24 for 650 GHz and 0.33 for 850 GHz. To estimate this cross/co-polar ratio for the wideband FTS measurement we take the average of the result for the three bands and as the error bar the standard deviation of the three measurements: $0.28\pm 0.05$.

\begin{figure*}
\includegraphics[width=0.4\textwidth]{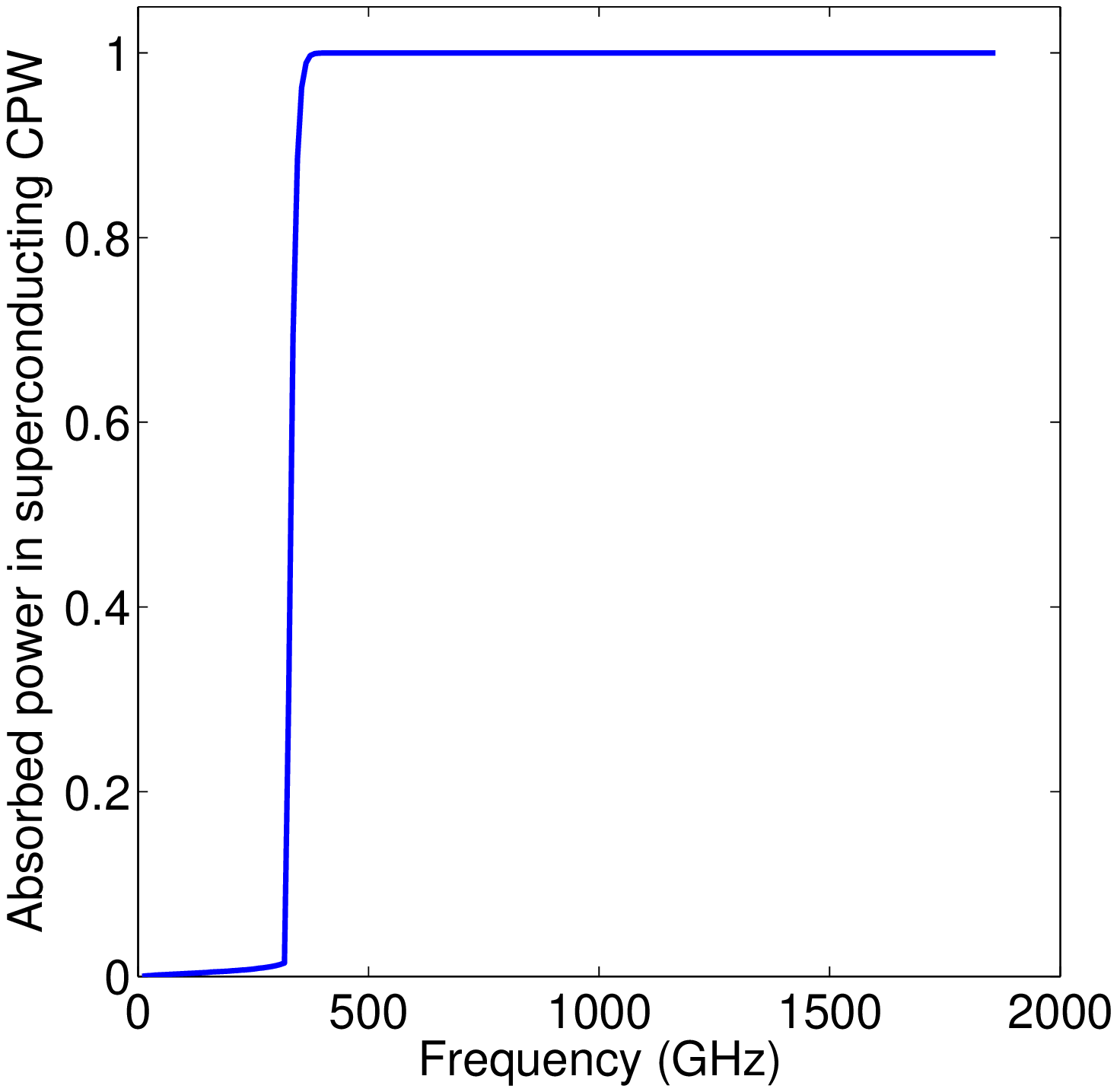}
\includegraphics[width=0.4\textwidth]{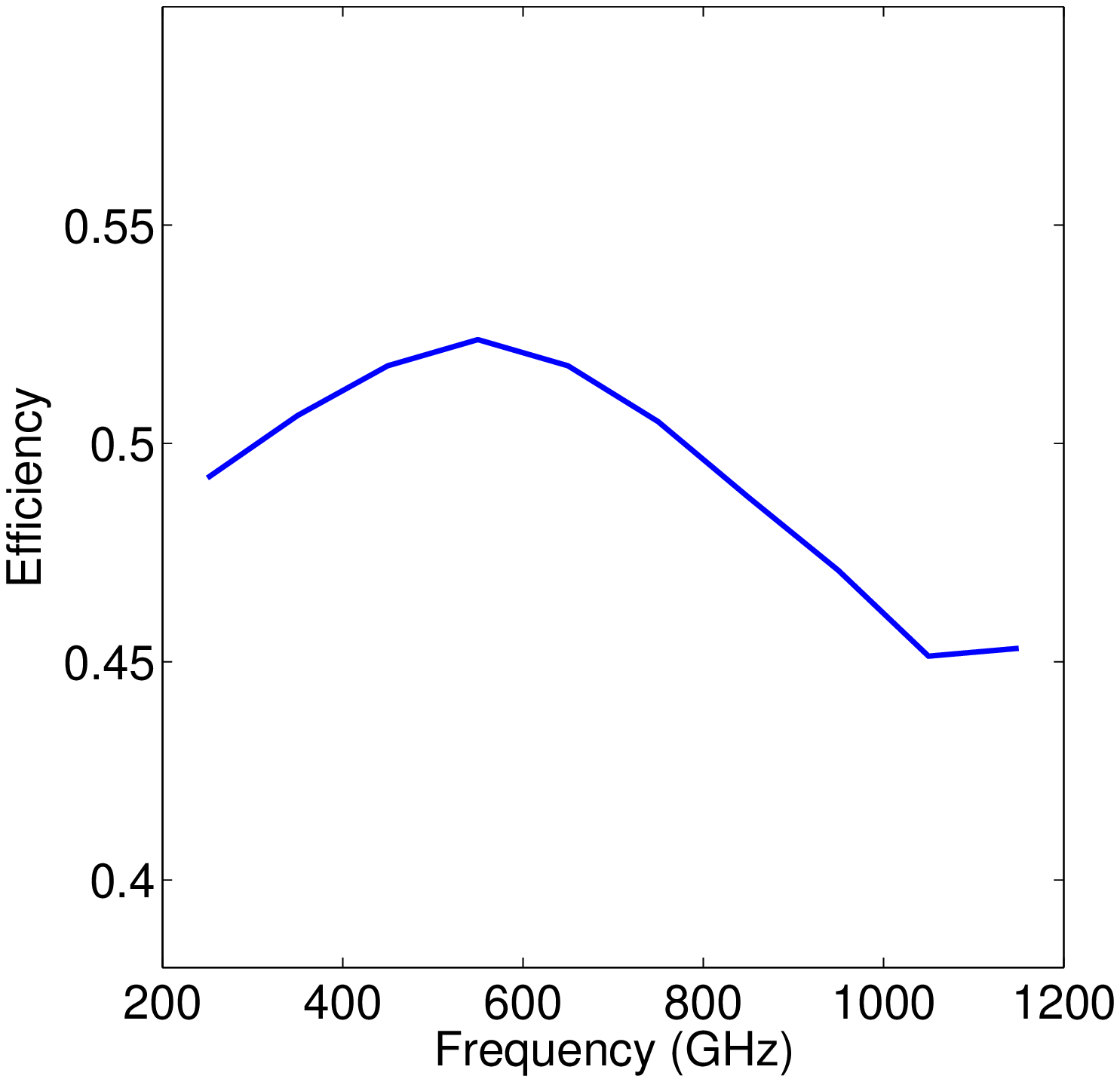}
\caption{\label{fig:efficiency} (Left) The absorbed power in the superconducting coplanar waveguide detector as a function of frequency and as a fraction of the power at the CPW input. (Right) The efficiency of the antenna from a CST simulation, including impedance mismatch of the antenna to CPW, front-to-back ratio, spillover and lens reflection. The multiplication of the results in these two panels is the simulated green curve in Fig. 2 in the main text.}
\end{figure*}

\end{document}